\documentclass[10pt]{IEEEtran15}
\usepackage{graphicx}
\usepackage{amsmath}
\usepackage{amssymb}
\usepackage{color}
\usepackage{epsfig}

\begin{document}
\twocolumn

\title{\LARGE Belief Propagation Based Decoding of Large Non-Orthogonal 
STBCs }
\author{Madhekar Suneel, Pritam Som, A. Chockalingam, and B. Sundar Rajan \\
\vspace{-0mm}
{\normalsize Department of ECE, Indian Institute of Science,
Bangalore 560012, INDIA} 
\vspace{-8mm}
}
\maketitle
\thispagestyle{empty}

\begin{abstract}
In this paper, we present a belief propagation (BP) based algorithm 
for decoding non-orthogonal space-time block codes (STBC) from cyclic 
division algebras (CDA) having {\em large dimensions}. The proposed 
approach involves message passing on Markov random field (MRF) 
representation of the STBC MIMO system. Adoption of BP approach 
to decode non-ortho\-gonal STBCs of large dimensions has not been 
reported so far. Our simulation results show that the proposed 
BP-based decoding achieves increasingly closer to SISO AWGN performance
for increased number of dimensions. In addition, it also achieves
near-capacity turbo coded BER performance; for e.g., with BP decoding
of $24\times 24$ STBC from CDA using BPSK (i.e., 576 real dimensions) 
and rate-1/2 turbo code (i.e., 12 bps/Hz spectral efficiency), coded 
BER performance close to within just about 2.5 dB from the theoretical 
MIMO capacity is achieved. 
\end{abstract} 

\vspace{-3.0mm}

\vspace{-5.0mm}
\section{Introduction}
\label{sec1}
\vspace{-2.5mm}
Use of multiple antennas at the transmitter can offer the benefits 
of transmit diversity (e.g., using space-time coding) and high data 
rates (e.g., using spatial multiplexing) \cite{jafarkhani}. MIMO 
systems that employ non-orthogonal space-time block codes (STBC) 
from cyclic division algebras (CDA) for arbitrary number of transmit 
antennas, $N_t$, are particularly attractive because they can 
simultaneously provide both {\em full-rate} (i.e., $N_t$ complex
symbols per channel use, which is same as in V-BLAST) as well as 
{\em full transmit diversity} \cite{bsr},\cite{cda}. The $2\times 2$ 
Golden code is a well known non-orthogonal STBC from CDA for 2 transmit 
antennas \cite{gold05}. High spectral efficiencies of the order of tens 
of bps/Hz can be achieved using large non-orthogonal STBCs. For e.g.,  
a $16\times 16$ STBC from CDA has 256 complex symbols in it with 512 real 
dimensions; with 16-QAM and rate-3/4 turbo code, this system offers a high 
spectral efficiency of 48 bps/Hz. Decoding of non-orthogonal STBCs with 
such large dimensions, however, has been a challenge. Sphere decoder and 
its low-complexity variants are prohibitively complex for decoding such 
STBCs with hundreds of dimensions. Recently, we proposed a low-complexity 
near-ML achieving algorithm to decode large non-orthogonal STBCs from CDA; 
this algorithm, which is based on  bit-flipping approach, is termed as 
likelihood ascent search (LAS) algorithm \cite{jsac}-\cite{gcom08}. Our 
new contribution in this paper is that we present a {\em belief propagation
(BP) based approach} to decoding of non-orthogonal STBCs with large 
dimensions, and report very good uncoded BER and near-capacity performance.
To our knowledge, BP has not been reported for decoding non-orthogonal
STBCs from CDA having large dimensions. 

Belief propagation \cite{merl} is known to be well suited in several 
communication problems \cite{frey}; e.g., decoding of turbo codes
and LDPC codes \cite{bp_turbo},\cite{ldpc}, multiuser detection 
\cite{bpmud1},\cite{bpmud2}, signal detection in ISI channels 
\cite{isi1},\cite{isi2}, and MIMO detection \cite{ieee06},\cite{icc07}.
Taking the cue from the success of BP in decoding turbo codes and
achieving near-capacity performance with large frame sizes, in this 
paper we achieve near-ML and near-capacity performance
in decoding STBCs with large dimensions using BP at practical 
complexities. It is known that graphical models that represent MIMO 
systems are highly connected. While BP was initially formalized for 
loop-free graphs, it has been empirically found to work in loopy graphs 
as well \cite{bp_turbo},\cite{loopy}. In the context of MIMO detection 
using BP, \cite{ieee06} reported a successful adoption of BP algorithm 
on Markov random fields (MRF) by employing belief propagation meant for 
pairwise MRFs as described in \cite{merl}. It presented the BER 
performance of the BP based detector for V-BLAST with $N_t=4$ and 
$N_r\hspace{-0.50mm}=\hspace{-0.50mm}4,6,8$ without and with simulated 
annealing. However, BP approach applied to large dimension decoding in 
MIMO systems, particularly large non-orthogonal STBC MIMO systems, has 
not been reported so far. In this regard, our work here is the first in 
reporting BP on MRFs for large dimension STBC MIMO decoding. Our simulation 
results show that the proposed BP decoding algorithm performs close to 
within just about 2.5 dB from the theoretical MIMO capacity in a 
$24\times 24$ non-orthogonal STBC MIMO system using BPSK and rate-1/2 
turbo code at a spectral efficiency of 12 bps/Hz. We also present the 
performance of BP decoding in the presence of spatial correlation. We 
show that the loss in performance due to spatial correlation can be 
alleviated by using increased receive spatial dimensions.

\vspace{-3.5mm}
\section{Non-Orthogonal STBC MIMO System Model}
\label{sec2}
\vspace{-1.5mm}
Consider a STBC MIMO system with multiple transmit and multiple receive
antennas. An $(n,p,k)$ STBC is represented by a matrix
{\small ${\bf X}_c \in {\mathbb C}^{n \times p}$}, where $n$ and $p$
denote the number of transmit antennas and number of time slots,
respectively, and $k$ denotes the number of complex data symbols sent
in one STBC matrix. The $(i,j)$th entry in ${\bf X}_c$ represents the
complex number transmitted from the $i$th transmit antenna in the $j$th
time slot. The rate of an STBC is $\frac{k}{p}$.
Let $N_r$ and $N_t=n$ denote the number of receive and transmit antennas,
respectively. Let ${\bf H}_c \in {\mathbb C}^{N_r\times N_t}$ denote the
channel gain matrix, where the $(i,j)$th entry in ${\bf H}_c$ is the
complex channel gain from the $j$th transmit antenna to the $i$th receive
antenna. We assume that the channel gains remain constant over one STBC
matrix duration. Assuming rich scattering, we model the entries of
${\bf H}_c$ as i.i.d $\mathcal C \mathcal N(0,1)$. The received space-time
signal matrix, ${\bf Y}_c \in {\mathbb C}^{N_r \times p}$, can be written as
\vspace{-1.5mm}
\begin{eqnarray}
\label{SystemModel}
{\bf Y}_c & = & {\bf H}_c{\bf X}_c + {\bf N}_c,
\end{eqnarray}

\vspace{-2.5mm}
where ${\bf N}_c \in {\mathbb C}^{N_r \times p}$ is the noise matrix at
the receiver and its entries are modeled as i.i.d
$\mathcal C \mathcal N\big(0,\sigma^2=\frac{N_tE_s}{\gamma}\big)$,
where $E_s$ is the average energy of the transmitted symbols, and
$\gamma$ is the average received SNR per receive antenna \cite{jafarkhani},
and the $(i,j)$th entry in ${\bf Y}_c$ is the received signal at the $i$th
receive antenna in the $j$th time slot. Consider linear dispersion STBCs, 
where ${\bf X}_c$ can be written in the form \cite{jafarkhani}
\vspace{-1.0mm}
\begin{eqnarray}
\label{SystemModelStbcLpx}
{\bf X}_c & = & \sum_{i = 1}^{k} x_c^{(i)} {\bf A}_c^{(i)},
\end{eqnarray}

\vspace{-3.5mm}
where $x_c^{(i)}$ is the $i$th complex data symbol, and
${\bf A}_c^{(i)} \in {\mathbb C}^{N_t \times p}$ is its 
weight matrix. The received signal model in (\ref{SystemModel}) 
can be written in an equivalent V-BLAST form as
\vspace{-1.0mm}
\begin{eqnarray}
\label{SystemModelvec2}
{\bf y}_c & = & \sum_{i=1}^{k} x_c^{(i)}\, (\widehat{{\bf H}}_c\, {\bf a}_c^{(i)}) + {\bf n}_c \,\,\, = \,\,\, \widetilde{{\bf H}}_c {\bf x}_c + {\bf n}_c,
\end{eqnarray}

\vspace{-3.0mm}
where
{\small ${\bf y}_c \in {\mathbb C}^{N_rp \times 1} = vec\,({\bf Y}_c)$},
{\small $\widehat{{\bf H}}_c \in {\mathbb C}^{N_rp \times N_tp} = ({\bf I} \otimes {\bf H}_c)$},
{\small ${\bf a}_c^{(i)} \in {\mathbb C}^{N_tp \times 1} = vec\,({\bf A}_c^{(i)})$}, 
{\small ${\bf n}_c \in {\mathbb C}^{N_rp \times 1} = vec\,({\bf N}_c)$}, 
{\small ${\bf x}_c \in {\mathbb C}^{k \times 1}$} 
whose $i$th entry is the data symbol $x_c^{(i)}$, and
{\small $\widetilde{{\bf H}}_c \in {\mathbb C}^{N_rp \times k}$}
whose $i$th column is
$\widehat{{\bf H}}_c \, {\bf a}_c^{(i)}$, $i=1,2,\cdots,k$. 
For notational simplicity, we drop the subscripts $c$ 
in (\ref{SystemModelvec2}) and write
\vspace{-5.75mm}
\begin{eqnarray}
\label{SystemModelII}
{\bf y} & = & {\bf H} {\bf x} + {\bf n},
\end{eqnarray}

\vspace{-2.5mm}
where 
${\bf x} = {\bf x}_c$, 
${\bf H} = {\widetilde {\bf H}}$, 
${\bf y} = {\bf y}_c$, and 
${\bf n} = {\bf n}_c$. 
We note that (\ref{SystemModelII}) can be viewed as an equivalent
V-BLAST representation of the non-orthogonal STBC MIMO system.
We assume that the channel coefficients are known at the receiver
but not at the transmitter.

\vspace{-4.00mm}
\subsection{High-Rate Non-Orthogonal STBCs from CDA}
\vspace{-2.5mm}
We consider square (i.e.,
$n\hspace{-0.9mm}=\hspace{-0.9mm}p\hspace{-0.9mm}=\hspace{-0.9mm}N_t$),
full-rate (i.e.,
$k\hspace{-0.9mm}=\hspace{-0.9mm}pn\hspace{-0.9mm}=\hspace{-0.9mm}N_t^2$),
circulant (where the weight matrices ${\bf A}_c^{(i)}$'s are
permutation type), non-orthogonal STBCs from CDA \cite{bsr}, whose
construction for arbitrary number of transmit antennas $n$ is given by
the matrix in (4.a) given at the bottom of this page \cite{bsr}.
In (4.a), {\small $\omega_n=e^{\frac{{\bf j}2\pi}{n}}$,
${\bf j}=\sqrt{-1}$, and $d_{u,v}$, $0\leq u,v \leq n-1$} are the $n^2$
data symbols from a QAM alphabet. When $\delta=t=1$, the code in (4.a) 
is information lossless (ILL), and when $\delta=e^{\sqrt{5}\,{\bf j}}$ and 
$t=e^{{\bf j}}$, it is of full-diversity and information lossless (FD-ILL) 
\cite{bsr}. High spectral efficiencies with large $N_t$ can be achieved 
using this code construction; e.g., $16\times 16$ STBC from (4.a) using 
4-QAM and rate-3/4 turbo code offers a spectral efficiency of 24 bps/Hz 
along with the full-diversity of order $N_tN_r$ under ML detection. 
However, since these STBCs are non-orthogonal, ML detection gets 
increasingly impractical for large $N_t$. Hence, a key challenge in 
realizing the benefits of these large non-orthogonal STBCs in practice 
is that of achieving near-ML performance for large $N_t$ at low decoding 
complexities. The BP based decoding approach in this paper essentially 
addresses this issue. 

\vspace{-2.5mm}
\section{BP on MRFs for Large Dimension MIMO 
Detection/Decoding }
\label{sec3}
\vspace{-2.0mm}
In this section, we present the BP detection algorithm assuming a V-BLAST 
system with BPSK modulation. The algorithm can be applied on the equivalent
V-BLAST representation of the non-orthogonal STBC MIMO system given by 
(\ref{SystemModelII}). BP is a technique that solves inference problems 
using graphical models such as factor graphs, Bayesian \hspace{0.5mm} 
belief \hspace{0.5mm} networks 

{\small
\thanks{
\line(1,0){505}
\[
\label{eqn}
\hspace{1.3cm}
\left[
\begin{array}{ccccc}
\sum_{i=0}^{n-1}d_{0,i}\,t^i & \delta\sum_{i=0}^{n-1}d_{n-1,i}\,\omega_n^i\,t^i & \delta\sum_{i=0}^{n-1}d_{n-2,i}\,\omega_n^{2i}\,t^i & \cdots & \delta\sum_{i=0}^{n-1}d_{1,i}\,\omega_n^{(n-1)i}\,t^i \\
\sum_{i=0}^{n-1}d_{1,i}\,t^i & \sum_{i=0}^{n-1}d_{0,i}\,\omega_n^i\,t^i & \delta\sum_{i=0}^{n-1}d_{n-1,i}\,\omega_n^{2i}\,t^i & \cdots & \delta\sum_{i=0}^{n-1}d_{2,i}\,\omega_n^{(n-1)i}\,t^i \\
\sum_{i=0}^{n-1}d_{2,i}\,t^i & \sum_{i=0}^{n-1}d_{1,i}\,\omega_n^i\,t^i & \sum_{i=0}^{n-1}d_{0,i}\,\omega_n^{2i}\,t^i & \cdots & \delta\sum_{i=0}^{n-1}d_{3,i}\,\omega_n^{(n-1)i}\,t^i \\
\vdots & \vdots & \vdots & \vdots & \vdots \\
\sum_{i=0}^{n-1}d_{n-2,i}\,t^i & \sum_{i=0}^{n-1}d_{n-3,i}\,\omega_n^i\,t^i & \sum_{i=0}^{n-1}d_{n-4,i}\,\omega_n^{2i}\,t^i & \cdots & \delta \sum_{i=0}^{n-1}d_{n-1,i}\,\omega_n^{(n-1)i}t^i \\
\sum_{i=0}^{n-1}d_{n-1,i}\,t^i & \sum_{i=0}^{n-1}d_{n-2,i}\,\omega_n^i\,t^i & \sum_{i=0}^{n-1}d_{n-3,i}\,\omega_n^{2i}\,t^i & \cdots & \sum_{i=0}^{n-1}d_{0,i}\,\omega_n^{(n-1)i}\,t^i
\end{array}
\right]. \hspace{10mm} (\mbox{4.a})
\]
}
}

and MRFs. In this paper, we consider BP on MRF 
representation of MIMO systems.

\vspace{-5.5mm}
\subsection{MRF Representation of a MIMO System}
\vspace{-2mm}
MRFs are graphs that indicate inter-dependencies between
random variables \cite{frey}. An MRF is an undirected graph whose 
vertices are random variables. The variables are such that any 
variable is independent of all the other variables, given its 
neighbors, i.e., 
\vspace{-2mm}
\begin{equation}
\text{p}\left(x_k\vert x_1,\ldots ,x_{k-1},x_{k+1},\ldots,x_N\right) = \text{p}\left[x_k\vert \mathcal{N}\left(x_k\right)\right],
\end{equation}

\vspace{-3mm}
where $\mathcal{N}\left(x_k\right)$ represents the set of all nodes
neighboring the node pertaining to the variable $x_k$.
Usually, the variables in an MRF are constrained by a \textit{compatibility
function}, also known as a \textit{clique potential} in literature.  A
\textit{clique} of an MRF is a fully connected sub-graph that does not
remain fully connected if any additional vertex of the MRF is included in it. 
This is sometimes called a \textit{maximal} clique, but we shall use the 
term clique to refer to a maximal clique. Let there be $N_C$ cliques
in the MRF, and $\mathbf{x}_j$ be the set of variables in clique $j$. Let
$\psi_j\left(\mathbf{x}_j\right)$ be the clique potential of clique $j$.  
Then the joint distribution of the variables is 
\vspace{-2mm}
\begin{equation}
\text{p}\left(\mathbf{x}\right) \propto \prod_{j = 1}^{N_C} \psi_j\left(\mathbf{x}_j\right),
\end{equation}

\begin{figure}
\centering
\includegraphics[width=0.21\textwidth]{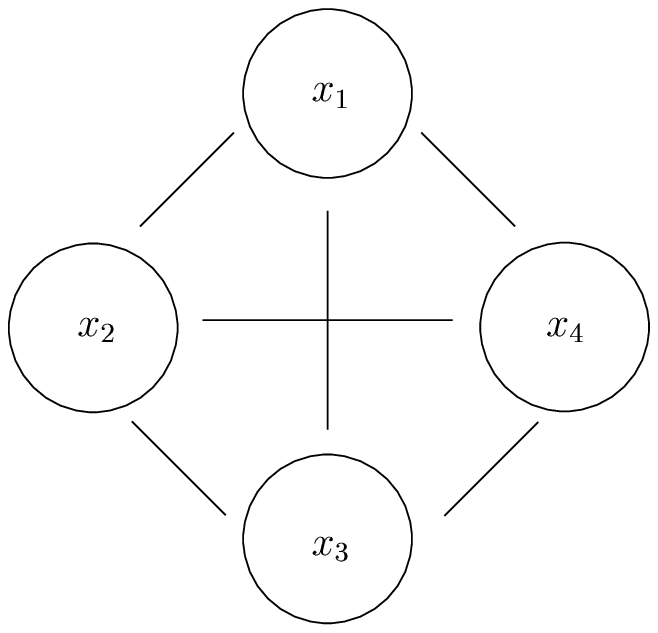}
\vspace{-2mm}
\caption{Markov random field representation of a BPSK V-BLAST 
system with 4 transmit antennas. }
\vspace{-10mm}
\label{figmrfexample}
\end{figure}

\vspace{-3mm}
where ${\bf x}$ is the set of all variables in the graph.
For example, consider the MRF representation of a BPSK V-BLAST system 
with 4 transmit antennas as shown in Fig. \ref{figmrfexample}. Here, 
$x_1, x_2, x_3, x_4$ respectively are the binary symbols transmitted 
from the four transmit antennas. Each of these symbols assumes a value 
from $\left\{\pm 1\right\}$. In a V-BLAST system, since every transmitted 
symbol interferes with every other transmitted symbol at the receiver, the 
MRF is fully-connected, and contains a single clique, namely, 
$\left\{x_1,x_2,x_3,x_4\right\}$. The joint probability distribution is 
\vspace{-3mm}
\begin{multline}
\text{p}\left(x_1,x_2,x_3,x_4\right) \,\, = \,\, \text{p}\left(x_1\right) \text{p}\left(x_2\vert x_1\right) \text{p}\left(x_3\vert x_1,x_2 \right) \\ \text{p}\left(x_4\vert x_1,x_2,x_3\right)\, .
\end{multline}

\vspace{-12mm}
\subsection{Pairwise MRFs}
\vspace{-0.75mm}
An MRF is called a \textit{pairwise} MRF if all the cliques in the 
\newpage
MRF are of size two. In this case, the clique potentials are all 
functions of two variables. The clique potentials can then be denoted as
$\psi_{i,j}\left(x_i,x_j\right)$, where $x_i,x_j$ are variables
connected by an edge in the MRF.

\begin{figure}
\centering
\includegraphics[width=0.2025\textwidth]{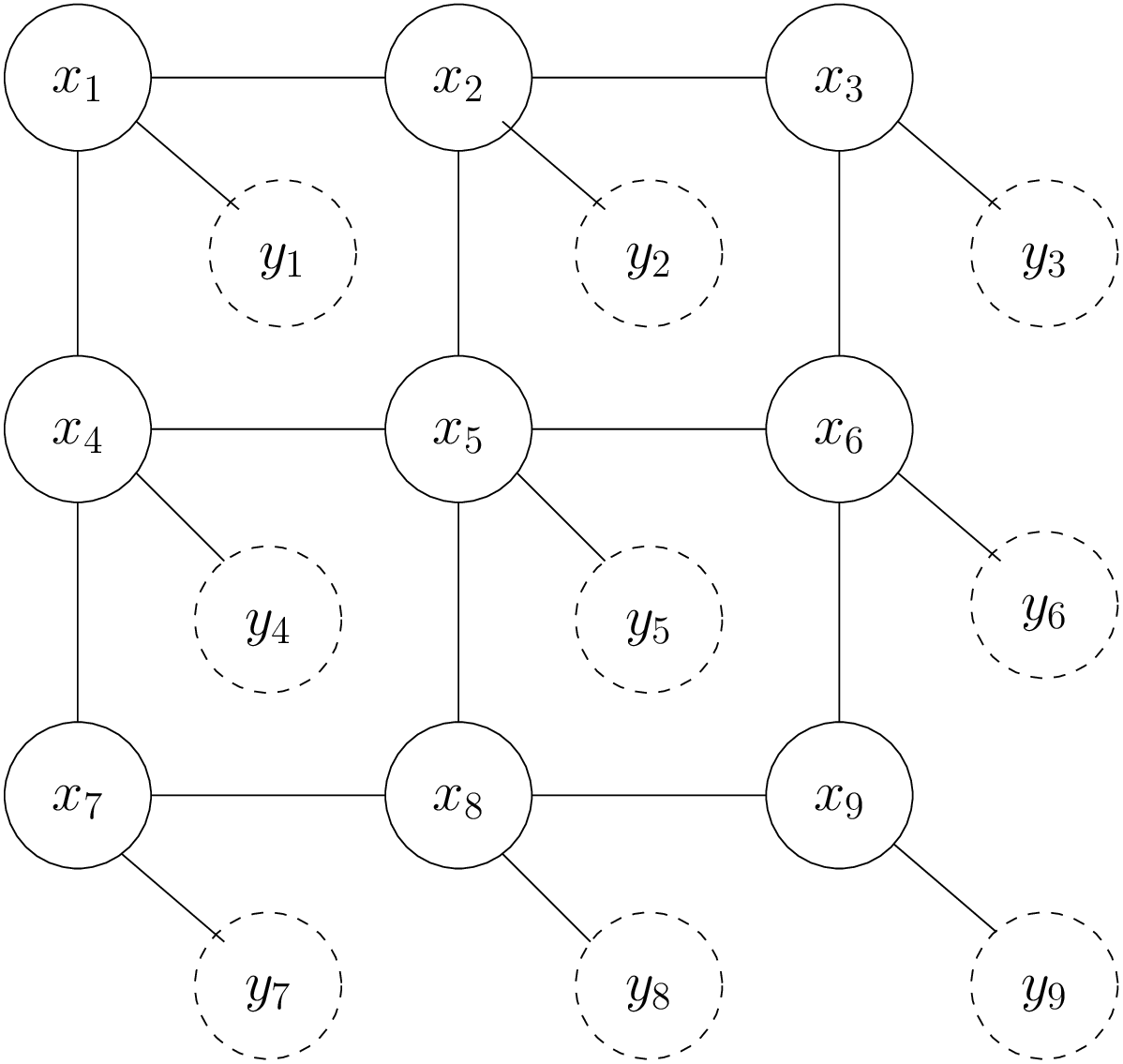}
\vspace{-2mm}
\caption{An example of a pairwise Markov random field with observed 
(explicit) variables and hidden (implicit) variables.}
\vspace{-5mm}
\label{figpairwisemrf}
\end{figure}

Consider a pairwise MRF in which the $x_i$'s denote underlying
\textit{hidden} variables on which the observed variables $y_i$'s
are dependent \cite{merl}. Let the dependence between the hidden variable 
$x_i$ and the explicit variable $y_i$ be represented by a \textit{joint}
compatibility function $\phi_{i}\left(x_i,y_i \right)$. This scenario
is shown in Fig. \ref{figpairwisemrf}. In such a scenario, the joint
distribution of the hidden and explicit variables is 
\vspace{-1mm}
\begin{equation}
\text{p}\left(\mathbf{x},\mathbf{y}\right) \propto \prod_{i,j}\psi_{i,j}\left(x_i,x_j\right) \, \prod_{i}\phi_i\left(x_i,y_i\right) \, .
\label{eqnpairwisemrf}
\end{equation}

\vspace{-10mm}
\subsection{Belief Propagation on Pairwise MRFs}
\vspace{-2mm}
Consider a situation similar to that of Fig. \ref{figpairwisemrf}, where
$x_i$'s are the hidden variables and $y_i$'s are the observed variables.
If we consider the $y_i$'s to be fixed and write $\phi_i\left(x_i,y_i\right)$
simply as $\phi_i\left(x_i\right)$, then, from (\ref{eqnpairwisemrf}), the
joint distribution for the hidden variables can be written as \cite{merl} 
\vspace{-1mm}
\begin{equation}
\text{p}\left(\mathbf{x}\right) \propto \prod_{i,j}\psi_{i,j} \left(x_i,x_j\right) \prod_i\phi_i\left(x_i\right) \, .
\label{eqnmrfjointdistribution}
\end{equation}

\vspace{-4mm}
A \textit{message} from node $j$ to node $i$ denoted as 
$m_{j,i}\left(x_i\right)$, and belief at node $i$ denoted as $b_i(x_i)$
are vectors of length equal to the number of values that the discrete
variable $x_i$ can possibly take (e.g., length of message and belief 
vectors is 2 in BPSK since $x_i \in \{\pm 1\}$). Each element of the 
belief vector is proportional to how likely the corresponding value of 
$x_i$ was transmitted. On the other hand, each element in the message 
vector $m_{ji}(x_i)$ is proportional to how likely $x_j$ thinks the 
corresponding value of $x_i$ has been transmitted.
The belief at node $i$ about the state of $x_i$ is 
\vspace{-2mm}
\begin{equation}
\text{b}_i\left(x_i\right) \propto \phi_i\left(x_i\right) \prod_{j \in \mathcal{N} \left(i\right)} m_{j,i}\left(x_i\right) \, .
\label{eqnmrfbelief}
\end{equation}

\vspace{-4mm}
In particular, the messages are defined as \cite{merl}
\vspace{-1mm}
\begin{equation}
\hspace{-0mm}
m_{j,i}\left(x_i\right) \propto \sum_{x_j}\phi_j\left(x_j\right) \psi_{j,i}\left(x_j,x_i\right) \prod_{k\in \mathcal{N}\left(j\right) \setminus i} \hspace{-2mm} m_{k,j}\left(x_j\right). \hspace{-4mm}
\label{eqnmsgdefn}
\end{equation}

\vspace{-3mm}
Equation (\ref{eqnmsgdefn}) actually constitutes an iteration, as the 
message is defined in terms of the other messages. Therefore, belief 
propagation essentially involves computing the outgoing messages from 
a node to each of its neighbors using the local joint-compatibility 
function and the incoming messages and transmitting them.

\begin{figure}
\centering
\includegraphics[width=0.475\textwidth]{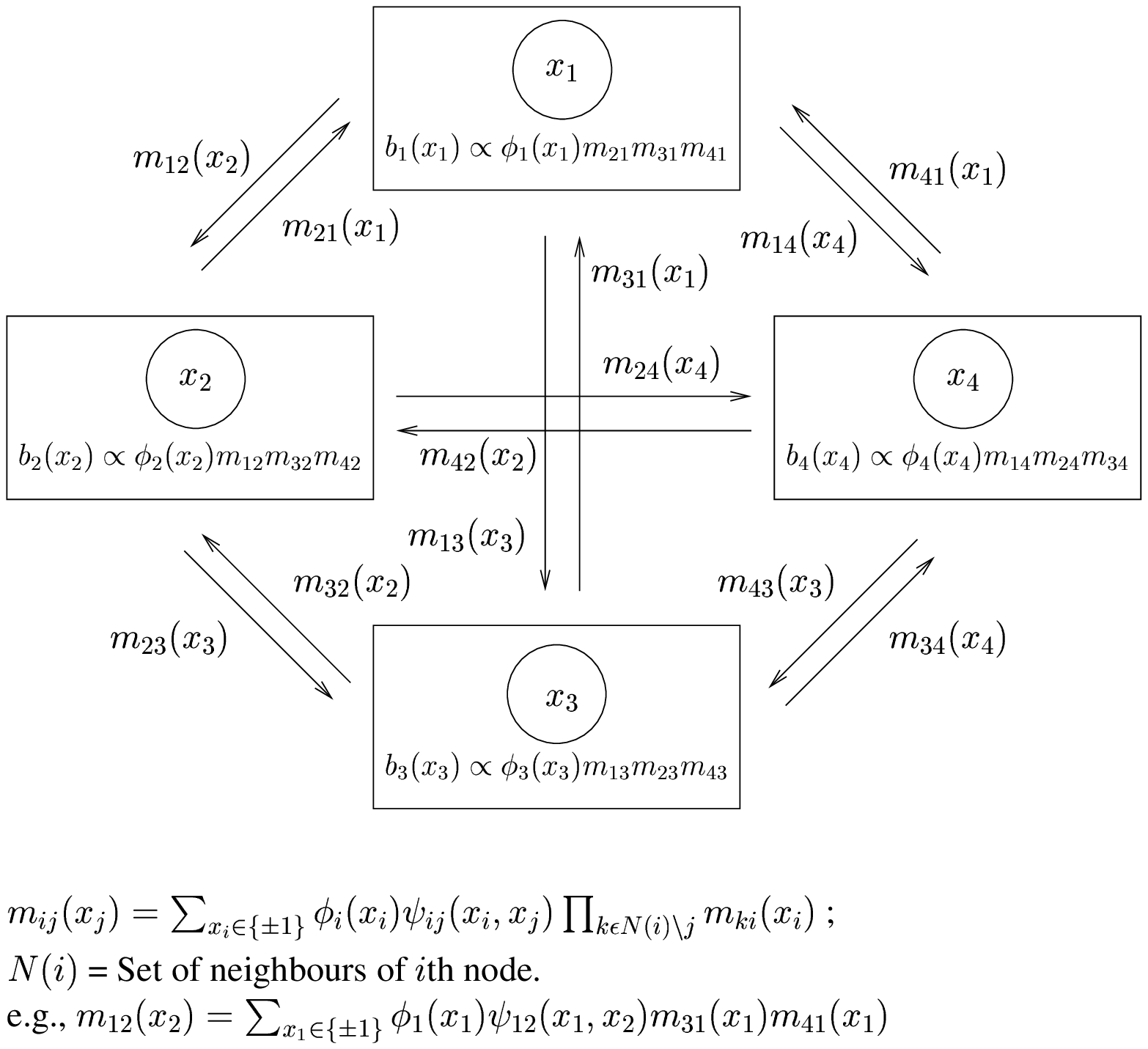}
\vspace{-1mm}
\caption{Message passing in an MRF for a BPSK V-BLAST system
with 4 transmit antennas.}
\vspace{-4mm}
\label{figmsg}
\end{figure}

\vspace{-4.0mm}
\subsection{BP for MIMO Detection}
\label{secb}
\vspace{-1.0mm}
In this subsection, for the MIMO system defined by an MRF (e.g., as 
shown in the graph in Fig. \ref{figmrfexample}), we present the BP 
based detection algorithm. We observe that the MRFs for MIMO systems, 
exemplified in Fig. \ref{figmrfexample}, are not pairwise MRFs. However, 
application of the BP algorithm for pairwise MRFs, as outlined by 
(\ref{eqnmrfbelief}) and (\ref{eqnmsgdefn}), to the MRFs of MIMO systems 
yields a low-complexity detector for MIMO systems with large 
dimensions\footnote{Our simulation results show that near SISO AWGN 
uncoded BER performance and near-capacity coded BER performance are 
achieved in large dimension V-BLAST as well as non-orthogonal STBC 
MIMO systems by this approximate approach.}. The joint probability 
distribution for hidden variables $x_i$ in a pairwise MRF is given 
by (\ref{eqnmrfjointdistribution}). We use the following functions 
$\phi_i$ and $\psi_{i,j}$ for MIMO detection using BP: 
\vspace{-1mm}
\begin{align}
\phi_i\left(x_i\right) & = \exp\left[ \Re\left(x_i^{H}z_i\right) + \text{ln}\left\{ \text{p}\left(x_i\right) \right\} \right] \, , \label{eqnoptphi} \\
\psi_{i,j}\left( x_i,x_j\right) & = \Re\left(\exp\left(- x_i^{H}R_{i,j}x_j \right)\right), 
\label{eqnoptpsi}
\end{align}
where \cite{ieee06}
\vspace{-4mm}
\begin{align}
\mathbf{z} & \triangleq \frac{1}{\sigma^2}\,\mathbf{H}^{H}\mathbf{y} \, ,\label{eqnz} \\
\mathbf{R} & \triangleq \frac{1}{\sigma^2}\,\mathbf{H}^{H}\mathbf{H} \label{eqnR} \, .
\end{align}

\vspace{-3mm}
Then, the message from node $i$ to a neighboring node $j$ in the BP 
algorithm is given by (\ref{eqnmsgdefn}) as given in \cite{merl}. For 
a BPSK V-BLAST system with 4 transmit antennas, the message passing 
scheme is summarized in Fig. \ref{figmsg}. Nodes pass messages to each 
other in an iterative fashion. After the last iteration, beliefs are 
calculated locally at each node. The value of $x_i$ which has the maximum 
belief is selected as the symbol decision. The belief is the soft-output 
of the detector which can be fed to the decoder in a coded system.

\vspace{-4mm}
\subsection{Computational Complexity}
\vspace{-2mm}
The complexity of the detection scheme comprises of three components, 
namely, $i)$ computation of channel correlation ${\bf R}$ given by
(\ref{eqnR}) and matched filter output ${\bf z}$ given by (\ref{eqnz}), 
$ii)$ computation of local evidence $\phi$ and compatibility function 
$\psi$ in (\ref{eqnoptphi}) and (\ref{eqnoptpsi}), respectively, and $iii)$ 
calculation of beliefs and messages during iterative message passing
given by (\ref{eqnmrfbelief}) and (\ref{eqnmsgdefn}), respectively. 
The computation of ${\bf R}$ and ${\bf z}$ involves the computation of 
${\bf H}^H{\bf H}$ and ${\bf H}^H{\bf y}$, respectively.
In case of STBC MIMO system, two good properties of the STBCs from CDA
are instrumental in achieving low orders of complexity for the computation 
of ${\bf H}^H{\bf H}$ and ${\bf H}^H{\bf y}$. They are: $i)$ the weight 
matrices ${\bf A}_c^{(i)}$'s are {\em permutation type}, and $ii)$ the 
$N_t^2\times N_t^2$ matrix formed with $N_t^2\times 1$-sized 
${\bf a}_c^{(i)}$ vectors as columns is a {\em scaled unitary matrix}.
For $N_t=N_r$, the computation of ${\bf H}^H{\bf y}$ and hence ${\bf z}$ 
can be done in $O(N_t^4)$ complexity, i.e., in $O(N_t^2)$ per-symbol
complexity since there are $N_t^2$ symbols in one STBC matrix. Likewise, 
the computation of ${\bf H}^H{\bf H}$ and hence ${\bf R}$ can be done in 
$O(N_t^3)$ per-symbol complexity. In case of V-BLAST, ${\bf R}$ and 
${\bf z}$ can be computed in $O(N_t^2)$ and $O(N_t)$ per-symbol 
complexity, respectively. Computation of $\psi$ involves $O(N_t^2)$ and 
$O(N_t)$ per-symbol complexity for STBC and V-BLAST, respectively. 
The per-symbol complexity of computing $\phi$ is $O(1)$. The per-symbol 
complexities involved in the computation of messages and beliefs in a 
single iteration for V-BLAST are $O(N_t^2)$ and $O(N_t)$, respectively; 
for STBC, these complexities are of order $O(p^2N_t^2)$ and $O(pN_t)$, 
respectively. 

\vspace{-4.5mm}
\section{Simulation Results }
\label{sec4}
\vspace{-2.0mm}
Our simulation results have shown that the proposed BP based algorithm
achieves increasingly closer to SISO AWGN uncoded BER performance for
V-BLAST signals with increasing number of dimensions (e.g., performance
close to within 1 dB of $10^{-3}$ uncoded BER for hundreds of dimensions). 
Since the dimensions in V-BLAST are in space alone, systems with hundreds 
of antennas may not be realistic. On the other hand, use of non-orthogonal 
STBCs from CDA can create hundreds of dimensions with just tens of antennas 
(space) and tens of channel uses (time). In this section, we present the
uncoded and coded BER performance of the proposed BP algorithm in decoding
large non-orthogonal STBCs. 5 BP iterations are used in all the simulations.

\vspace{-4.5mm}
\subsection{Uncoded BER performance of large STBCs from CDA:}
\begin{figure}
\hspace{-3mm}
\includegraphics[width=0.525\textwidth,height=0.295\textheight]{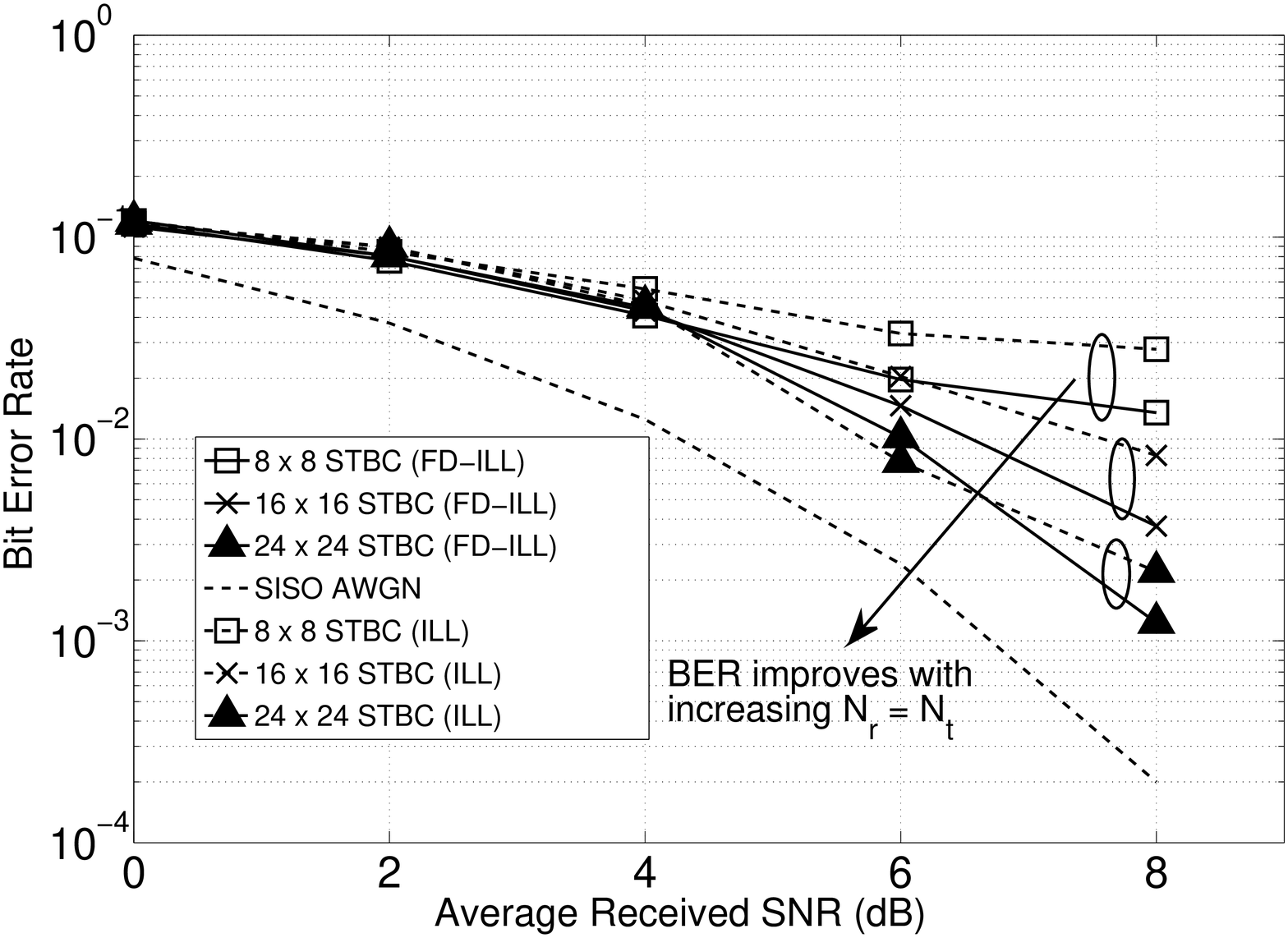}
\vspace{-8mm}
\caption{Uncoded BER of BP decoding of $8\times 8$, $16\times 16$ 
and $24\times 24$ non-orthogonal STBCs. ILL ($\delta=t=1$) and FD-ILL 
({\scriptsize $\delta=e^{\sqrt{5}{\bf j}}, t=e^{{\bf j}}$}) STBCs. 
$N_t=N_r$, BPSK, 5 BP iterations. {\em BP decoding achieves near SISO 
AWGN performance for large sized STBCs. }}
\vspace{-5mm}
\label{fig_uncoded_stbc}
\end{figure}

\vspace{-2.0mm}
{\em BP decoding achieves near SISO AWGN performance for large STBCs:}
In Fig. \ref{fig_uncoded_stbc}, we plot the uncoded BER performance
of $8\times 8$ (64 dimensions), $16\times 16$ (256 dimensions) and 
$24\times 24$ (576 dimensions) non-orthogonal STBCs from CDA for BPSK 
and $N_t=N_r$, as a function of average received SNR per receive antenna,
$\gamma$ \cite{jafarkhani}. BER plots for STBCs with ILL (i.e., 
$\delta=t=1$) and FD-ILL (i.e., $\delta=e^{\sqrt{5}{\bf j}}, \, 
t=e^{{\bf j}}$) are shown. For reference purposes, we have plotted the 
BPSK BER performance on a SISO AWGN channel as well as on a SISO 
flat-fading channel. From Fig. \ref{fig_uncoded_stbc}, the following
two interesting observations can be made: 
\begin{itemize}
\vspace{-1mm}
\item BERs of both ILL and FD-ILL STBCs with BP decoding improve 
      and approach SISO AWGN performance as the number of dimensions 
      (i.e., $N_t^2$) is increased. For e.g., the performance of 
      $24\times 24$ FD-ILL STBC is just about 1.5 dB away from 
      SISO AWGN performance at $10^{-3}$ BER. This is due to the 
      inherent ability of BP to perform well in large systems. 
\item With the proposed BP decoding, the BER of ILL STBC is worse 
      than the BER of FD-ILL STBC. This performance gap between ILL 
      and FD-ILL STBCs diminishes with increasing $N_t$, indicating 
      that for large $N_t=N_r$ ILL feature of the STBC with $\delta=t=1$
      is good enough.
\end{itemize}

\begin{figure}
\centering
\includegraphics[width=0.525\textwidth,height=0.295\textheight]{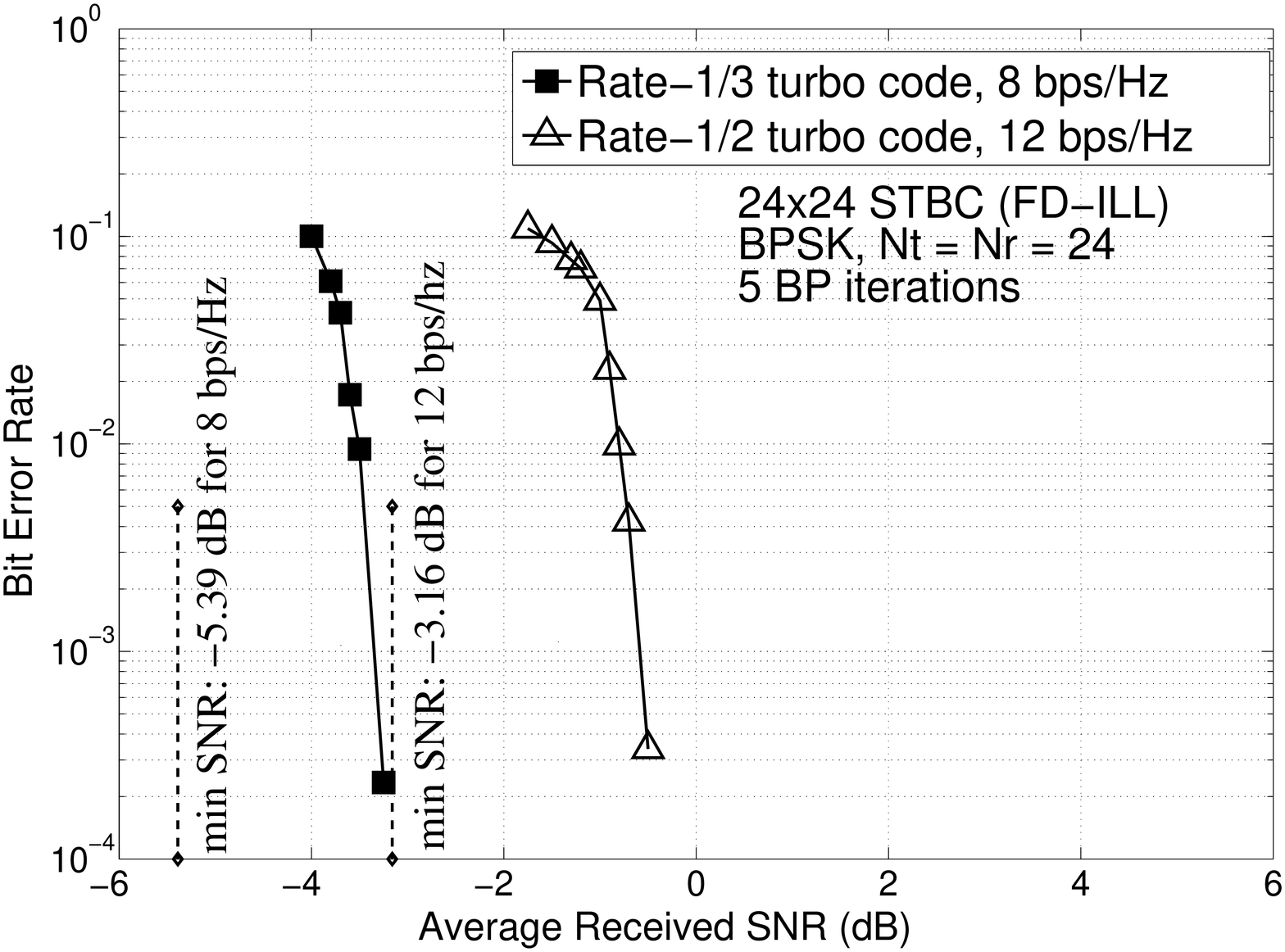}
\vspace{-8mm}
\caption{Turbo coded BER of BP decoding of $24\times 24$ non-orthogonal 
FD-ILL STBC ({\scriptsize $\delta=e^{\sqrt{5}{\bf j}}, t=e^{{\bf j}}$}). 
$N_t=N_r=24$, BPSK, turbo code rates: rate-1/2 (12 bps/Hz) and rate-1/3 
(8 bps/Hz). 5 BP iterations. {\em BP decoding achieves near-capacity 
performance to within about 2.5 dB from capacity}. }
\vspace{-3mm}
\label{fig_coded_stbc}
\end{figure}

{\em Turbo coded BER performance:} 
Figure \ref{fig_coded_stbc} shows the turbo coded BER performance of 
the BP detector in a STBC MIMO system with BPSK and $24\times 24$ 
FD-ILL STBC (i.e., $\delta=e^{\sqrt{5}{\bf j}}$, $t=e^{{\bf j}}$) 
and $N_t=N_r=24$.  Rate-1/3 and rate-1/2 turbo codes achieving 
8 bps/Hz and 12 bps/Hz spectral efficiencies, respectively, are used. 
The theoretical minimum SNRs required to achieve these 8 and 12 bps/Hz 
capacities in a $N_t=N_r=24$ MIMO channel, as computed from the ergodic 
capacity formula \cite{jafarkhani}, are also shown in Fig. 
\ref{fig_coded_stbc}. From Fig. \ref{fig_coded_stbc}, it can be seen  
that the vertical fall in coded BER occurs only about 2.5 dB away from 
the theoretical minimum SNRs, which is very good in terms of nearness 
to capacity. Such nearness results for large STBCs from CDA using BP 
decoding have not been reported so far.

\begin{figure}
\centering
\includegraphics[width=0.525\textwidth,height=0.295\textheight]{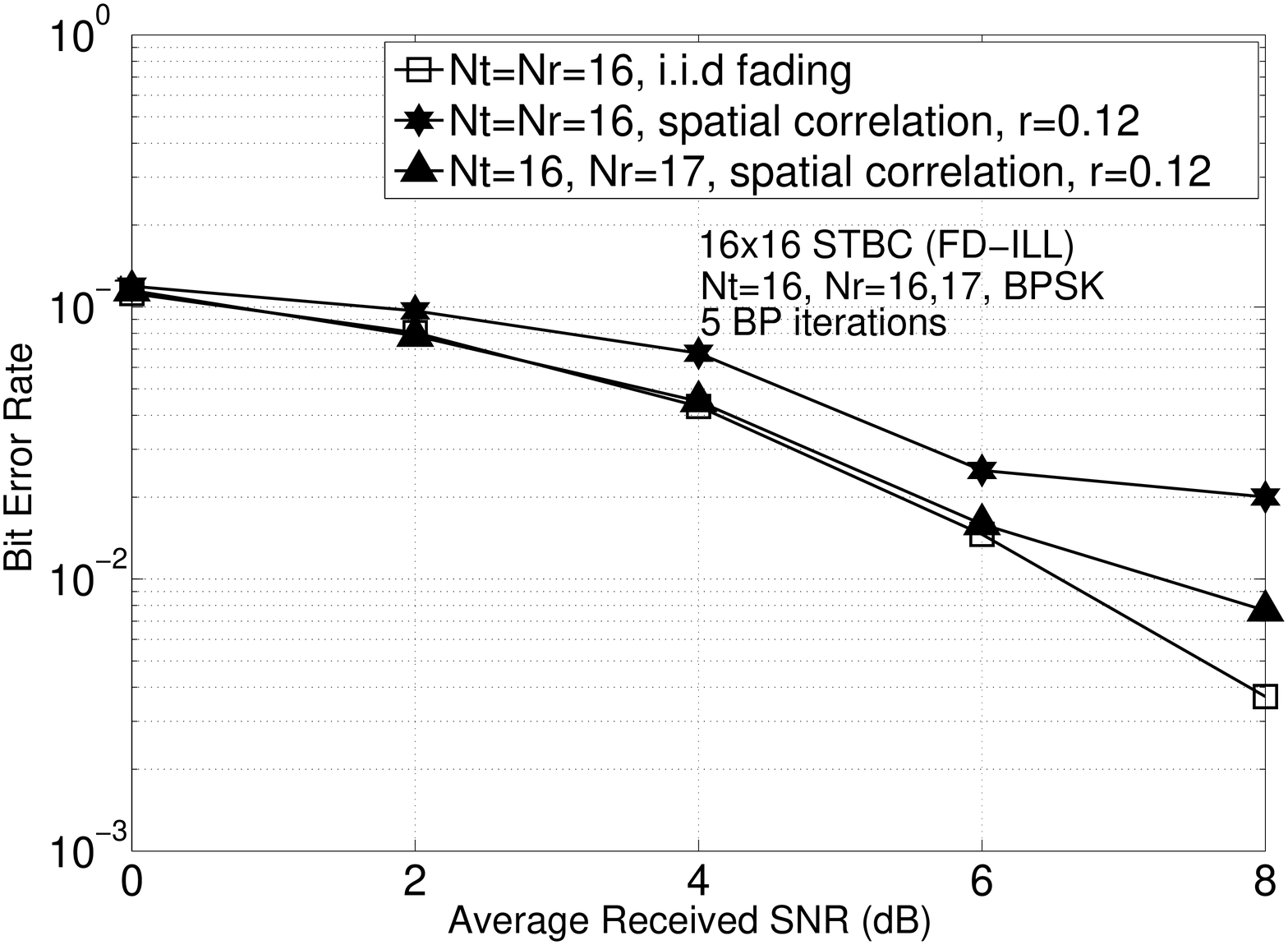}
\vspace{-8mm}
\caption{Effect of spatial correlation on the uncoded BER of BP 
decoding of $16\times 16$ non-orthogonal FD-ILL STBC 
({\scriptsize $\delta=e^{\sqrt{5}{\bf j}}, t=e^{{\bf j}}$}). 
$N_t=16$, $N_r=16,17$, BPSK, 5 BP iterations, $r=0.12$. {\em Spatial 
correlation degrades achieved diversity order compared to that 
achieved in i.i.d fading. 
Increasing $N_r$ alleviates this performance loss.} }
\vspace{-4mm}
\label{fig_spatial}
\end{figure}

{\em Effect of Spatial Correlation:}
In generating the BER results in Figs. \ref{fig_uncoded_stbc} and 
\ref{fig_coded_stbc}, we have assumed i.i.d. fading. However, MIMO 
propagation conditions witnessed in practice often render the i.i.d. 
fading model as inadequate. More realistic MIMO channel models that 
take into account the scattering environment, spatial correlation, 
etc., have been investigated in the literature \cite{mimo1}-\cite{corr2}. 
For example, spatial correlation at the transmit and/or receive 
side can affect the rank structure of the MIMO channel resulting in 
degraded MIMO capacity \cite{mimo1}. The structure of scattering in 
the propagation environment can also affect the capacity \cite{mimo2}. 
Hence, it is of interest to investigate the performance of the proposed
BP decoder in more realistic MIMO channel models. Towards this end,
in this subsection, we adopt the correlated MIMO channel model in 
\cite{corr2}, which incorporates the single spatial correlation parameter,
$r$, presented in \cite{corr1}, to a matrix channel model. Figure
\ref{fig_spatial} shows the simulated uncoded BER performance of 
$16\times 16$ FD-ILL STBC with BPSK, $N_t=16$, $N_r=16,17$ for the 
correlation channel model in \cite{corr2} with $r=0.12$. Performance
of the same with i.i.d fading and $N_t=N_r=16$ is also plotted for
comparison. From Fig. \ref{fig_spatial}, it can be observed that
compared to i.i.d fading, there is a loss in diversity order in 
spatial correlation for {\small $N_t=N_r=16$}; further, use of more 
receive antennas ({\small $N_r=17, N_t=16$}) alleviates this loss in 
performance. We note that the proposed BP based decoding can be used 
to decode perfect codes \cite{perf06},\cite{perf07} of large dimensions 
as well.

\vspace{-4mm}
\section{Conclusion}
\label{sec5}
\vspace{-1mm}
We presented a low-complexity decoding scheme based on BP to decode 
non-orthogonal STBCs from CDA having large dimensions. The proposed
BP scheme involved message passing on Markov random field representation 
of the {\small STBC MIMO} system. Successful application of BP for detection 
in large non-orthogonal STBC MIMO systems has not been reported so far. 
Simulation results showed that the BP approach to large STBC decoding 
is quite effective, achieving near-ML and near-capacity performance
in STBCs with large dimensions. Effect of spatial correlation on the 
performance of the proposed BP decoding was presented. Extension of the 
proposed BP approach to higher order modulation (e.g.,$M$-PAM/$M$-QAM) 
in the large STBC MIMO context can be investigated as further extension 
to this work.


\vspace{-3.0mm}
{\footnotesize
\begin{thebibliography}{99}
\vspace{-0mm}
\bibitem{jafarkhani}
H. Jafarkhani, {\em Space-Time Coding: Theory and Practice}, Cambridge
University Press, 2005.

\bibitem{bsr}
B. A. Sethuraman, B. Sundar Rajan, and V. Shashidhar, ``Full-diversity
high-rate space-time block codes from division algebras,'' {\em IEEE
Trans. Inform. Theory}, vol. 49, no. 10, pp. 2596-2616, October 2003.

\bibitem{cda}
F. Oggier, J.-C. Belfiore, and E. Viterbo, {\em Cyclic Division Algebras:
A Tool for Space-Time Coding,} Foundations and Trends in Commun. and
Inform. Theory, vol. 4, no. 1, pp. 1-95, Now Publishers, 2007.

\bibitem{gold05}
J.-C. Belfiore, G. Rekaya, and E. Viterbo, ``The golden code: A $2\times 2$
full-rate space-time code with non-vanishing determinants,'' {\em IEEE
Trans. Inform. Theory}, vol. 51, no. 4, April 2005.

\bibitem{jsac}
K. Vishnu Vardhan, Saif K. Mohammed, A. Chockalingam, B. Sundar Rajan,
``A low-complexity detector for large MIMO systems and multicarrier CDMA
systems,'' {\em IEEE JSAC Spl. Iss. on Multiuser Detection, for Adv. Commun.
Systems and Networks}, 
pp. 473-485, April 2008. 

\bibitem{isit08}
Saif K. Mohammed, A. Chockalingam, and B. Sundar Rajan, ``A low-complexity 
near-ML performance achieving algorithm for large MIMO detection,'' 
{\em Proc. IEEE ISIT'2008}, Toronto, July 2008. 

\bibitem{gcom08}
Saif K. Mohammed, A. Chockalingam, and B. Sundar Rajan, ``High-rate
space-time coded large MIMO systems: Low-complexity detection and
performance,'' {\em Proc. IEEE GLOBECOM'2008}, December 2008.

\bibitem{merl}
J. S. Yedidia, W. T. Freeman, Y. Weiss, ``Understanding belief
propagation and its generalizations,'' {\em MERL Tech Rep. TR-2001-22},
Jan. 2002.

\bibitem{frey}
B. J. Frey, {\em Graphical Models for Machine Learning and Digital 
Communication,} Cambridge: MIT Press, 1998.

\bibitem{bp_turbo}
R. J. McEliece and D. J. C. MacKay, and J-F. Cheng, ``Turbo decoding
as an instance of Pearl's belief propagation algorithm,'' {\em IEEE Jl. 
Sel. Areas in Commun.,} vol. 16, no.2, pp. 140-152, February 1998.

\bibitem{ldpc}
D. J. C. MacKay, ``Good error-correcting codes based on very sparse
matrices,'' {\em IEEE Trans. Inform. Theory}, 
pp. 399-431, March 1999.

\bibitem{bpmud1}
A. Montanari, B. Prabhakar, and D. Tse, ``Belief propagation based
multiuser detection,'' Online arXiv:cs/0510044v2 [cs.IT] 22 May 2006.

\bibitem{bpmud2}
D. Guo and C-C. Wang, ``Multiuser detection of sparsely spread CDMA,''
{\em IEEE JSAC Spl. Iss. on Multiuser Detection, for Adv. Commun.
Systems and Networks}, vol. 26, no. 3, pp. 421-431, April 2008.

\bibitem{isi1}
O. Shental, A. J. Weiss, N. Shental, Y. Weiss, ``Generalized 
belief propagation receiver for near-optimal detection of two-dimensional
channels with memory,'' {\em IEEE Inform. Theory Workshop}, 
October 2004.

\bibitem{isi2}
G. Colavolpe and G. Germi, ``On the application of factor graphs and the
sum-product algorithm to ISI channels,'' {\em IEEE Trans. on Commun.,}
vol. 53, no. 5, pp. 818-825, May 2005.

\bibitem{ieee06}
J. Soler-Garrido, R. J. Piechocki, K. Maharatna, and D. McNamara,
``Analog MIMO detection on the basis of belief propagation,'' 
{\em Proc. IEEE Mid-West Symp. on Circuits and Systems, 2006}.

\bibitem{icc07}
X. Yang, Y. Xiong, F. Wang, ``An adaptive MIMO system based on unified
belief propagation detection,'' {\em Proc. IEEE ICC'2007}, June 2007.

\bibitem{loopy}
K. Murphy, Y. Weiss, and M. Jordan, ``Loopy belief propagation for approximate
inference: An empirical study,'' {\em 15th Annual Conf. on Uncertainty
in Artificial Intelligence}, pp. 467-470, 1999.

\bibitem{mimo1}
D. Shiu, G. J. Foschini, M. J. Gans, J. M. Khan, ``Fading correlation
and its effect on the capacity of multi-antenna systems,'' {\em IEEE Trans.
Commun.,} vol. 48, pp. 502-513, March 2000.

\bibitem{mimo2}
D. Gesbert, H. B\"olcskei, D. A. Gore, A. J. Paulraj, ``Outdoor
MIMO wireless channels: Models and performance prediction,'' {\em
IEEE Trans. on Commun.,} vol. 50, pp. 1926-1934, December 2002.

\bibitem{corr2}
A. van Zelst and J.S. Hammerschmidt, ``A single coefficient spatial 
correlation model for multiple-input multiple-output (MIMO) radio 
channels,'' {\em 27th General Assembly of the Intl. Union of Radio 
Science (URSI)}, Maastricht, Netherlands, August 2002.

\bibitem{corr1}
G. D. Durgin and T. S. Rappaport, ``Effects of multipath angular 
spread on the spatial correlation of received voltage envelopes,''
{\em Proc. Proc. IEEE VTC'1999},  pp. 996-1000, 1999.

\bibitem{perf06}
F. E. Oggier, G. Rekaya, J.-C. Belfiore, and E. Viterbo, ``Perfect
space-time block codes,'' {\em IEEE Trans. on Inform. Theory},
vol. 52, no. 9, September 2006.

\bibitem{perf07}
P. Elia, B. A. Sethuraman, and P. V. Kumar, ``Perfect space-time codes for
any number of antennas,'' {\em IEEE Trans. Inform. Theory,}
vol. 53, no. 11, pp. 3853-3868, November 2007.

\end{thebibliography}			 
}
\end{document}